\newcommand\sun{\hbox{$\odot$}}

\documentclass{iau}
\usepackage{graphicx}

\title[Quenching in Simulations] 
{Red Galaxies from Hot Halos in Cosmological Hydro Simulations}

\author[Jared Gabor]   
{Jared Gabor$^1$
}

\affiliation{$^1$CEA Saclay \\
Bat. 709, Gif-sur-Yvette, 91191 France \\
email: {\tt jared.gabor@cea.fr}}

\pubyear{2013}
\volume{295}  
\pagerange{xx--xx}
\jname{The intriguing life of massive galaxies}
\editors{D. Thomas, A. Pasquali \& I. Ferreras, eds.}
\begin{document}

\maketitle

\begin{abstract}
I highlight three results from cosmological hydrodynamic simulations that
yield a realistic red sequence of galaxies: 1) Major galaxy mergers
are not responsible for shutting off star-formation and forming the
red sequence.  Starvation in hot halos is.  2) Massive galaxies grow
substantially ($\sim \times 2$ in mass) after being quenched,
primarily via minor (1:5) mergers. 3) Hot halo quenching naturally
explains why galaxies are red when they either (a) are massive or (b)
live in dense environments.

\keywords{galaxies: evolution, galaxies: interactions}
\end{abstract}

\firstsection 
\section{Major mergers are not responsible for quenching}

\begin{figure}[]
\begin{center}
 \includegraphics[width=5in]{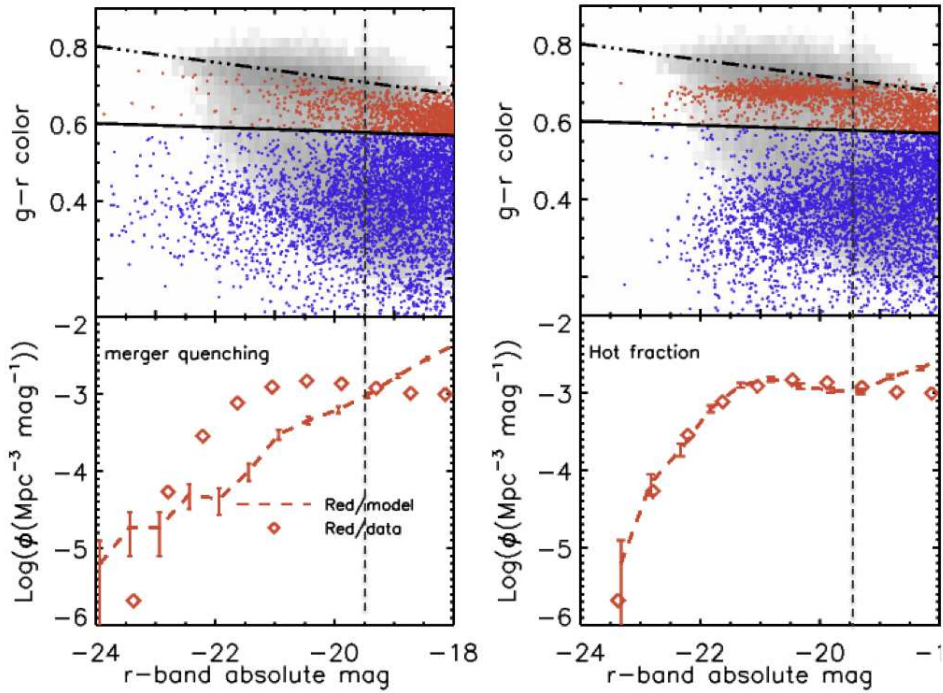} 
 \caption{Merger quenching (left) does not produce enough red
   galaxies, but hot gas quenching (right) does.  In color-magnitude
   diagrams (top panels) colored points represent simulated galaxies,
   and the grayscale represents SDSS galaxies.  In the luminosity
   functions (bottom panels), dashed lines with error bars show the
   number densities of simulated red galaxies, and diamonds
   those from SDSS.  Only hot halo quenching matches the $z\sim0$
   number densities of red galaxies.  Based on \cite[Gabor et
     al. 2011]{gabor11}.}
   \label{fig.cmdlf}
\end{center}
\end{figure}

The physical mechanism(s) responsible for shutting down star-formation
and leading to the red sequence are still debated (\cite[Bower et
  al. 2006]{bower06}, \cite[Croton et al. 2006]{croton06},
\cite[Hopkins et al. 2008]{hopkins08}, \cite[Somerville et
  al. 2008]{somerville08}).  Two leading pictures have emerged: a)
major mergers trigger starbursts and possibly AGNs, whose combined gas
consumption and feedback can rid the galaxy of gas to fuel
star-formation; and b) hot gas coronae form in massive halos,
shock-heating any infalling gas, and some heating process (such as a
radio AGN) prevents that gas from cooling.  Often, these are
respectively called ``quasar'' mode and ``radio'' mode AGN feedback,
alluding to the possible importance of AGN.

I have independently tested simplistic versions of these two
mechanisms in cosmological hydrodynamic simulations.  For merger
quenching, I identify mergers on-the-fly during the simulation and
eject all the gas from remnants in a 1000 km s$^{-1}$ wind.  For hot
halo quenching, I identify galaxies whose halos are dominated by hot
gas ($T>10^{5.4}$~K; \cite[Kere{\v s} et al. 2005]{keres05}), and
continuously add thermal energy to the circum-galactic gas around
them.  The results of these simulations are shown in Figure
\ref{fig.cmdlf} (cf. \cite[Gabor et al. 2011]{gabor11}) as color-magnitude diagrams
and luminosity functions.  Merger quenching fails to yield a
significant red sequence at $z=0$, whereas hot halo quenching forms
red galaxies in numbers consistent with observations.

In the context of cosmological models, major mergers are neither
necessary nor sufficient to explain the red sequence.  They are not
\emph{sufficient} because galaxies are constantly accreting new gas
from the cosmic web, even after mergers.  Since the accreted gas provides fuel for
star-formation, this accretion must be stopped to ensure that galaxies
become red and stay red.  Mergers are not \emph{necessary} because an
alternative quenching mechanism -- hot halo quenching  -- appears to
produce enough red galaxies.  Note that there are enough major mergers
to explain the numbers of red galaxies, if only the galaxies stopped
accreting after the merger \cite[(Gabor et al. 2010)]{gabor10}.

\section{Massive galaxies grow substantially after quenching}

\begin{figure}[]
\begin{center}
 \includegraphics[width=5in]{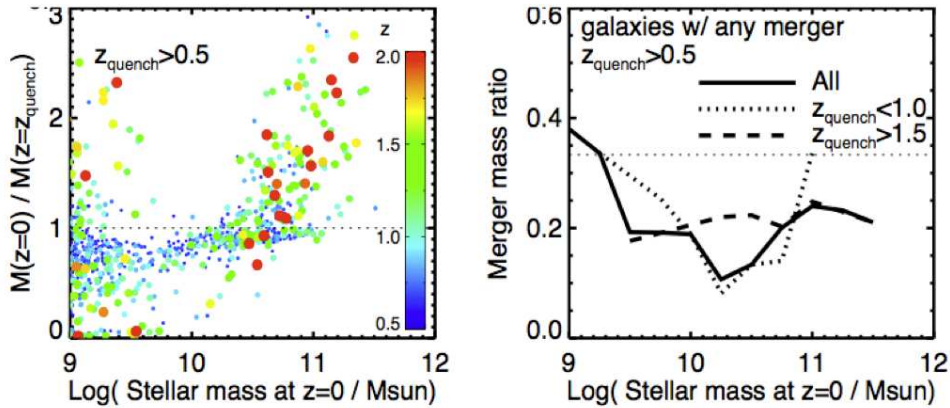} 
 \caption{After becoming red, galaxies grow significantly via minor
   mergers.  {\bf Left:} Mass growth of galaxies quenched at $z>0.5$,
   expressed as the $z=0$ stellar mass divided by the mass at the time
   of quenching.  Galaxies are color-coded by the redshift at which
   they were quenched.  Massive galaxies have typically grown by
   factors $\sim 2$.  {\bf
     Right:} Mass-weighted mean merger mass-ratio for galaxies
   quenched at $z>0.5$.  Dashed line at 0.33 shows the typical
   division between major and minor mergers (1:3).  Quenched galaxies
   grow mostly via minor ($\sim$ 1:5) mergers.  Taken from \cite[Gabor \& Dav\'e 2012]{gabor12}.}
   \label{fig.mass_growth}
\end{center}
\end{figure}

In the hot halo quenching simulation, massive galaxies typically turn
red at $z>0.5$, and grow significantly in mass after being quenched.
The mass growth is shown in the left panel of Figure
\ref{fig.mass_growth}.  Since star-formation is negligible in these
galaxies, there are only two ways to change in stellar mass: mass loss
from stellar evolution, and galaxy mergers.  By number, most galaxies
do not change much in mass -- these are mostly recently-quenched
satellites which are unlikely to merger with other satellites.
Massive galaxies, on the other hand, grow by factors up to 3, with a
large scatter driven by variations in merger history.

These massive galaxies are typically central galaxies accreting their
small satellites in minor mergers.  The right-hand panel of Figure
\ref{fig.mass_growth} shows the mean merger mass ratio (for those
galaxies with at least one merger) as a function $z=0$ stellar mass.
Here I have weighted each merger event by the mass of the smaller
galaxy.  Thus the plot shows the mass ratio which has been most
important for adding mass to galaxies in each bin of stellar mass.  In
all cases, the characteristic merger is below the typical 1:3 major
merger threshold, with a typical value of 1:5.  Minor mergers dominate
the mass growth of quenched galaxies.  This result implies strong
growth in the \emph{sizes} of quenched galaxies at high redshift
(\cite[Gabor et al. 2012]{gabor12}, \cite[Oser et al. 2012]{oser12}).


\section{Hot gas quenching explains both ``mass quenching'' and ``environment quenching''}

\begin{figure}[t]
\begin{center}
 \includegraphics[width=4.5in]{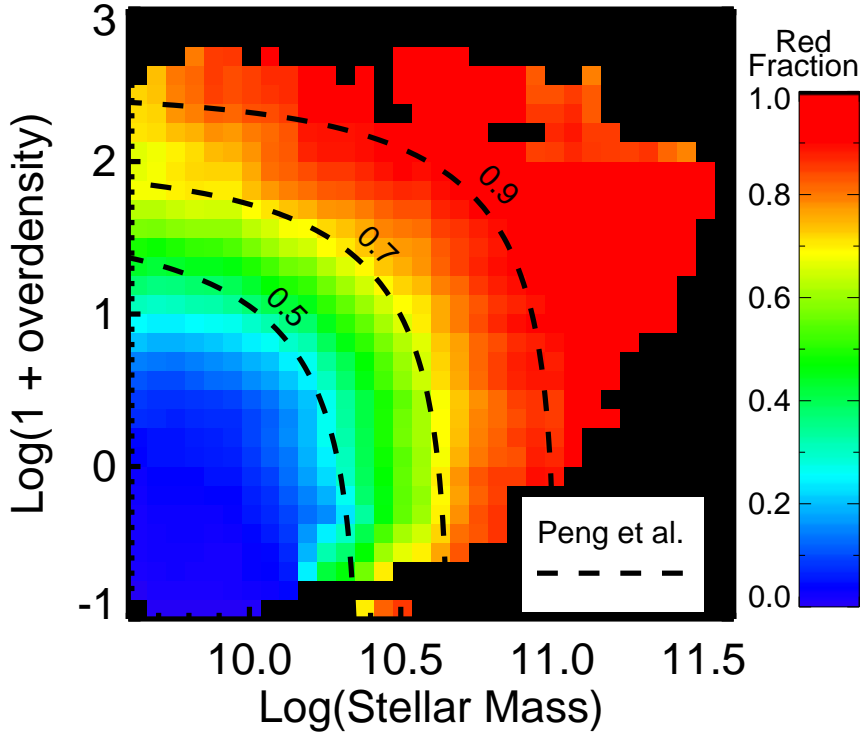} 
 \caption{Galaxies are quenched at high stellar masses and high
   overdensities.  I show the red fraction (color-coding) of galaxies
   as a function of stellar mass and local overdensity, a measure of
   environment.  Dashed lines are from the model of
   \cite[Peng et al. 2010]{peng10}.  The boxy contour shape suggests
   that ``environment quenching'' and ``mass quenching'' are
   independent, when in fact they both result from the presence of hot
   gas.  Taken from \cite[Gabor \& Dav\'e 2012]{gabor12}.}
   \label{fig.fred}
\end{center}
\end{figure}

In the hot halo quenching model, any galaxies that live in a hot corona are
starved of incoming fuel for star-formation.  A galaxy has two
alternative paths to live inside a hot halo: a) it is the central
galaxy in a halo of $> 10^{12} M_{\sun}$, where a hot coronae is
likely to form; or b) it is a satellite galaxy in such a halo.  Case
(a) can be thought of as ``mass quenching'' or ``central quenching'',
and case (b) can be thought of as ``environment quenching'' or
``satellite quenching.''

These two modes of quenching are apparent in Figure \ref{fig.fred},
inspired by \cite[Peng et al. (2010)]{peng10} and taken from
\cite[Gabor \& Dav\'e (2012)]{gabor12}.  The fraction of red galaxies
increases with overdensity (i.e. in denser environments) and with
stellar mass.  Moreover, the ``boxy'' shape of the contours suggests
that these modes are independent.  In our simulation, they both result
from hot gas cutting off the fuel supply for star-formation.

``Central quenching'' and ``satellite quenching'' can be explained
naturally in this model.  In hydrodynamic simulations, halos above
$\sim 10^{12} M_{\sun}$ are all dominated by hot gas (\cite[Birnboim \&
  Dekel 2003]{birnboim03}, \cite[Kere{\v s} et al. 2005]{keres05},
\cite[Gabor et al. 2010]{gabor10}).  Furthermore, in the absence of
quenching, stellar mass closely tracks halo mass.  So a star-forming
galaxy will increase its stellar mass as its halo mass increases due
to accretion.  Then, when the halo reaches $\sim 10^{12} M_{\sun}$
(i.e. stellar mass reaches $\sim 10^{10.5} M_{\sun}$), a hot corona
will form which quenches star-formation.  This manifests as a strong
increase in red galaxy fraction at stellar masses $\gtrsim 10^{10.5}
M_{\sun}$ -- mass quenching.

Once a massive galaxy is quenched in this way, its satellite galaxies
will also be quenched since they live in the same hot halo.  The dark
matter halo will continue to grow as additional galaxies fall in.
Such infalling galaxies typically start out as star-forming centrals,
but after becoming satellites they will be quenched by the hot gas
halo.  This is satellite quenching.  Satellites become quenched
regardless of their masses, as long as they live in sufficiently dense
environments where hot gas dominates.

\section{Summary}
Despite its simplicity, the hot halo quenching model does a remarkable
job of matching basic observables.  Although hot halos do not tell the
whole story, they appear to be the dominant factor in forming the red
sequence.  Mergers must play some role in the formation of today's red
ellipticals, but halting the inflow of new gas is crucial.

\end{document}